# Deployment of a Free-Text Analytics Platform at a UK National Health Service Research Hospital: CogStack at University College London Hospitals

Kawsar Noor[1,3], Lukasz Roguski[1,3], Alex Handy[1,3], Roma Klapaukh[1], Amos Folarin[2], Luis Romao[3], Joshua Matteson[3,4], Nathan Lea[1,3], Leilei Zhu[3] , Wai Keong Wong[3] ,Anoop Shah[1,3], Richard J Dobson[1,2]

## Abstract.

As more healthcare organisations transition to using electronic health record (EHR) systems it is important for these organisations to maximise the secondary use of their data to support service improvement and clinical research. These organisations will find it challenging to have systems which can mine information from the unstructured data fields in the record (clinical notes, letters etc) and more practically have such systems interact with all of the hospitals data systems (legacy and current). To tackle this problem at University College London Hospitals, we have deployed an enhanced version of the CogStack platform; an information retrieval platform with natural language processing capabilities which we have configured to process the hospital's existing and legacy records. The platform has improved data ingestion capabilities as well as better tools for natural language processing. To date we have processed over 18 million records and the insights produced from CogStack have informed a number of clinical research use cases at the hospitals.

## Section 1: Introduction

---

[1] University College London, Institute of Health Informatics
[2] King's College London
[3] University College London Hospitals NHS Foundation Trust, Clinical Research Informatics Unit
[4] Epic Systems Corporation

Over the past twenty years we have seen an increased uptake of electronic health records (EHR) within healthcare organisations with much of this being attributable to national efforts in having healthcare organisations transition to using full EHR systems [1] [2]. These EHRs represent a rich data asset but there remains a challenge in secondary use of the data for improving clinical care through activities such as service improvement and clinical research. In many cases the EHRs have simply replicated the paper system that they replaced and have not taken full advantage of the opportunities presented in having the health records in this new electronic format. Whilst functional systems to address these gaps are emerging many of the tools and data analytic approaches used on EHR data are limited to structured data, such as coded diagnoses and numeric clinical measurements, despite almost 80% of information being recorded as unstructured free text [10] (such as clinical notes, imaging reports and transfer of care documents). An additional difficulty is that a hospital's record is typically distributed across numerous disconnected data systems which presents a challenge in data harmonisation.

Working with EHRs thus presents a challenge firstly in harmonising and accessing the hospitals entire record from both existing and legacy data systems and secondly having tools and techniques available to mine and extract data from within these records; especially the unstructured free text. Manual analysis of unstructured text is time-consuming, so there has been much interest in developing automated methods for extracting accurate structured information from the text. Interpreting free text is a major analytic challenge; clinical text is written in a variety of styles by numerous authors, and may have mis-spellings, negations and information which does not relate to the patient. There has been intense interest in developing natural language processing techniques to interpret clinical text. Early methods used a rule-based approach, but more modern algorithms incorporate machine learning techniques, enabling the algorithms to 'learn' as more data is analysed.

The CogStack platform [6] was developed to address these exact problems. The platform can be described as an information retrieval (IR) system designed to interface with a hospital's EHR system. It was initially developed with an emphasis on ingestion and harmonisation of records from multiple data systems within a healthcare organisation. Whilst certain off the shelf natural language processing (NLP) tools were explored in the first iteration they were added as a proof of concept to demonstrate that the platform could potentially be configured to interact with such tools. In this paper we discuss the deployment of CogStack at University College London Hospital (UCLH) and the improvements introduced to the platform to ensure that platform scales to meet the growing research demands of the hospital. Our deployment has focused on addressing the following three key issues which need to be universally addressed at all research driven healthcare organisations.

**Multiple Data Systems:** The EHRs of an organisation will typically be distributed across a number of different vendor systems, posing a challenge for the use of this information for clinical care and research. It is not uncommon for an organisation to have to maintain oversight over a myriad of data systems and vendors due to the fact that different clinical specialties will have different requirements of how data needs to be stored and managed. The resulting heterogeneity in data means that it is challenging for the organisation to find a common data model or even process through which the organisation's entire record can be harmonised. Methods and systems through which data is stored, collected and retrieved have been improving in order to tackle this challenge. Most notably many NHS trusts have opted to transition into using full scale EHR systems (e.g. Epic) each of which typically enforce their own data models. Some systems such as Epic go further in providing additional systems that allow data from third party data and legacy systems to be integrated with data collected via their own systems (Epic Clarity/Caboodle). Messaging standards (e.g. HL7 FHIR[5]), standardised terminologies (e.g SNOMED CT ) and standardised clinical information models (such as openEHR archetypes[6]) aim to improve interoperability between systems, but much more work is needed in this area. In order to maximise the benefit of patient data, it is essential that clinicians and researchers can access data in a way that is flexible, easily adaptable and independent of the organisation's choice of current and previous EHR systems.

**Multiple Data Formats:** A patient's record may be distributed across both scanned documents (pdfs), text documents (.doc files) as well as data stored in relational databases. Legacy documents for example will likely be stored as files and attachments whereas data that has been generated using a modern EHR system will likely be stored in a more structured way; possibly in a relational database. An IR system would thus need to be able to ingest and interact with records from all the various data formats used by the organisation. The CogStack platform provides a distributed architecture for document processing, including PDF to text conversion, or optical character recognition that may be needed prior to analysis of the text itself.

**Unstructured Text:** A final issue is that data within the EHR systems is recorded in both structured and unstructured fields. Some information is inherently unstructured in nature and needs to be recorded as free text (e.g. patient stories), but even where structured fields are available, clinicians may not use them and enter the information in free text instead. For example, a recent audit in our Trust found that patients admitted with suspected or confirmed COVID-19 had only 62.3% of their key diagnoses and comorbidities recorded in the structured problem list [11]. In order to support use of clinical data at scale and for multiple stakeholders, a successful IR system should provide mechanisms through which the clinical information within the unstructured free text notes can be made available. The CogStack platform provides a convenient user interface for searching free text and invoking information extraction algorithms, presenting the results in a way that is easy to visualise and harness for downstream research or for reintegration as structured data back into the EHR.

---

[5] https://www.hl7.org/fhir/overview.html
[6] https://www.openehr.org/

# Section 2: Methods:

In this section we describe our deployment of CogStack Nifi[7], a text analytics and IR system that can ingest, harmonise and mine structured clinical data from both structured unstructured clinical text, and that we have deployed at UCLH hospitals.

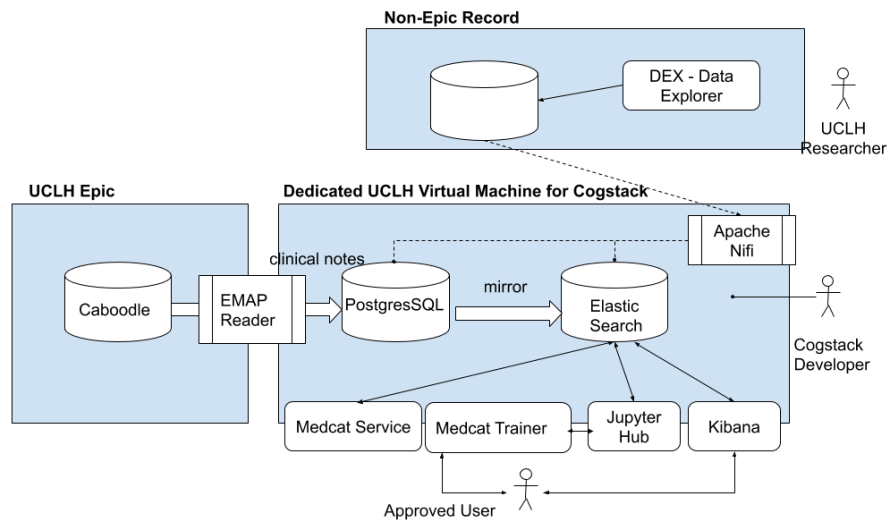

**Fig 1: An overview of the CogStack platform**

*Infrastructure*

The CogStack-Nifi platform builds upon the previous version [6] and has been deployed at multiple hospitals including South London and The Maudsley, Guy's and St Thomas', King's College Hospital and UCLH. This new iteration improves upon the previous version primarily in two aspects. The first is improved tools for clinicians to interact with and train clinical natural language processing systems and the second is the addition of an Apache-Nifi module for facilitating and managing data flows.

In the previous iteration the platform required a number of custom extract transform and load (ETL) scripts for managing the ingestion of data from the live record and legacy systems. This approach however does not scale well in practice and it can quickly become burdensome for developers to manage the various ETL scripts when the number of dataflows increases. Observing these difficulties, the Apache-Nifi module was added to the CogStack platform. Apache-Nifi is a visual interface for managing complex data flows between different data systems. Data flows in Apache-Nifi are depicted as directed graphs and provide useful visual feedback for system administrators such as the status of a particular data flow and number of documents processed. Most importantly Apache-Nifi is compatible with various data systems which means that the administrator is capable of writing the various ETL components in practically whatever programming language they choose. In addition as all of the dataflows are accessible within the interface it means maintaining oversight of all of the dataflows is considerably easier than having to monitor multiple custom ETL scripts. UCLH has in particular developed a number of Nifi workflows that are designed to work with the UCLH data warehouse as well as many legacy data systems. These components conduct the various extractions and data transformations necessary for downstream CogStack services. We discuss the various dataflows in the next section.

---

[7] https://github.com/CogStack/CogStack-NiFi

The CogStack-Nifi platform, as in the previous iteration, uses ElasticSearch (ES) for data storage. The Elasticsearch database is a 'no SQL database' that is a popular choice for systems that want to index data that is rich in text. It has been successfully deployed in various domains including e-commerce, legal systems and healthcare. In addition to ElasticSearch we have chosen to stage data from our EHR research database into a PostgresQL database so as to ensure we have a secondary backup of our data. In addition, the CogStack NLP models which are used to process the records upon ingestion also store the outputs within both the ES database as well as the PostgresQL database; this is useful as other research systems within the trust have the option of reading from either database.

Data Security and Governance

Use of unstructured EHR data for clinical research is challenging because of confidentiality concerns, leading to difficulty in obtaining ethics and information governance approvals for accessing such data. The CogStack approach is to embed text analytic capabilities and research staff within NHS Trusts, allowing sensitive text to be analysed in situ.
Although data is routinely ingested into the CogStack platform researchers wishing to use the data still need to undergo an approvals process to grant before accessing the data and/or making use of ML models trained on patient data for their research. UCLH has in place a system called Data Explorer[8] (DEX) through which researchers can apply for access to use clinical data. If researchers require CogStack a successful DEX application needs to be submitted and approved and the appropriate data protection impact assessment (DPIAs) completed. Each DPIA is assessed and approved by UCLH's information governance lead before the user is able to access the data on the CogStack platform and eventually the permission to process and analyse the data using CogStack trained ML models.

---

[8] https://www.uclhospitals.brc.nihr.ac.uk/clinical-research-informatics-unit/data-explorer

As data ingested into the CogStack platform contains patient sensitive data, all UCLH CogStack services are hosted within a secure environment that is only accessible within the hospital network. CogStack has a number of virtual machines that have been provisioned to process the trust's data. We have followed best practice for software deployment and have designated these virtual machines for development, testing and production.

In addition we have in place processes to remove patient identifiable data from the free text records before being used for research. CogStack has a de-identification module which is used to prepare batches of data for specific users, and can be deployed before or after ingestion into CogStack's central, standardised data lake. The module builds on the open source Philter library developed by the University of California, San Francisco which achieved over 99% recall on the benchmark I2B2 de-identification dataset by using a combination of rules based and statistical approaches [9]. In the following we detail the ingestion pipeline as well as how data is then accessed and processed once it has been ingested into ES.

## Data Ingestion:

CogStack uses Apache Nifi for managing data flows from the hospital's various data sources into CogStack's central database (and Elasticsearch instance). Using Apache Nifi we are able to define how the extract, transform and load (ETL) processes are implemented for each data source. We are able to set up dataflows that run periodically as well as manage ingestions that only happen once. Below we describe the various data sources from which we ingest from.

**Trust Data**: In 2019 UCLH officially transitioned to using Epic[9] as its primary EHR system. Prior to this the trust had a number of data systems for each of its departments/clinics. The Epic system has in place a number of databases that capture integrated hospital data. Its data warehouse, Caboodle, has been extended to capture non-EHR and historic data records as well. In particular UCLH has deployed the Epic Caboodle data warehouse for this purpose, and this is the primary database that CogStack ingests data from.

As this data is stored as relational data, setting up data flows into CogStack requires only that CogStack understands the data schema of the target database. The dataflows are then set up using Apache-Nifi as a batch process. The batch process transforms the data into a format that is compatible with Elasticsearch ingestion api. Most clinical research projects requiring CogStack to date have been retrospective studies and have not required access to a live data feed. Consequently the batch process runs on a daily basis and this can be easily modified as needed through the Nifi interface.

**Archived Data and Other Records:** A number of records in the Trust (such as those created prior to transition to Epic) are not included in Epic Caboodle data, and have required custom dataflows to be set up. Records in the legacy systems are often stored as documents that have been scanned as images or as text documents (.doc, .pdfs etc). In such cases CogStack uses Apache Tika's character recognition (OCR) in order to convert the contents of these documents into text that can then be persisted into our ES index.

---

[9] https://www.epic.com/

| Document Type | No of notes |
|---|---|
| Clinical Notes | 2,500,000 |
| Imaging Reports | 250,000 |
| Letters | 16,000,000 |

Table 1: Number of notes ingested and analysed by CogStack

**One-off-ingestions:** Whilst CogStack's primary focus is ingesting and processing data from the Trust, there are occasionally requests to analyse non-trust datasets. Examples of this include allergy reports taken from the National Reporting and Learning System. We describe this dataset in the case studies section. In such cases CogStack can accommodate these ad hoc requests via custom ingestion scripts using Apache Nifi.

## Data Storage:

Following ingestion, data is indexed into CogStack's Elasticsearch database. Elasticsearch (ES) is a No-SQL open source database. Underlying ES's technology is an open source search engine which provides an API for storing and retrieving documents using a custom query language called Lucene. As documents are ingested into the database, ES scores each doc ES computes various scores for each document using its NLP engine. These scores are later used to retrieve documents in a much faster and scalable way.

## Text Analytics:

Data stored within ES is stored as free-text. CogStack provides a number of tools which users can use to interact with data. Some of these tools are task specific and other tools allow one to programmatically extract data from the ES instance before running more advanced text-analytics/NLP models on the data.

## Kibana

For use cases where querying via keywords and other easy to define features and regular expressions is enough, CogStack uses the Kibana interface. The Kibana interface provides a view of the data that has been ingested into the ES index. Such searches are not easily handled by SQL and are more. These features enable users to construct large compound queries. In addition to its search functionality Kibana also provides some basic visualisation tools including.

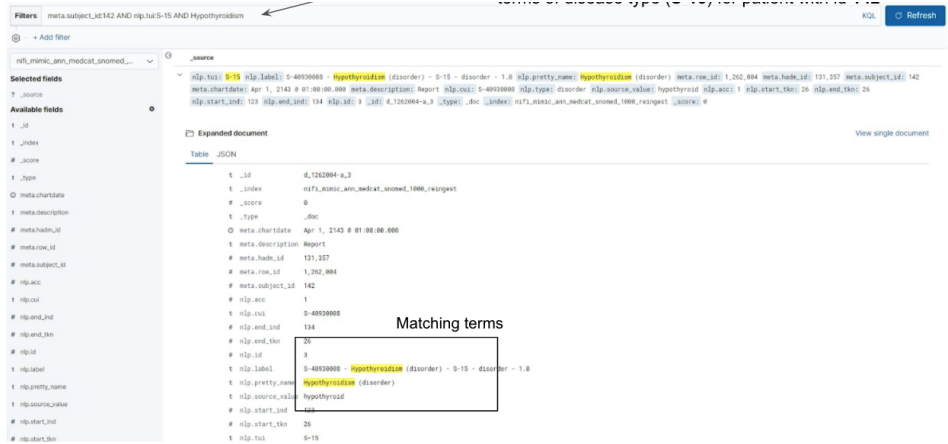

Fig 2: The Kibana interface being used to conduct keyword searches

However often in clinical research projects the requirements extend beyond keyword searches (information retrieval). A common requirement for many projects is to be able to identify the mention of certain clinical concepts (e.g. SNOMED CT/UMLS terms) in free text. Consider for example a Cardiologist who would like to know exactly what symptoms have been recorded about his patients in their clinical notes. In this case a tool is required that can identify and extract these terms from the free text. Our CogStack deployment uses a number of machine learning models that are capable of annotating the unstructured text and provides clinical labels that the user can use to better structure their queries.

## Machine Learning:

### MedCAT:

The machine learning model that identifies clinical concepts in free text is MedCAT [12]. Medcat is a named entity recognition and linking (NER+L) model that both identifies the clinical concepts in free text and then links them to a predefined medical ontology (e.g. SNOMED CT, UMLS). Medcat uses a dictionary based approach which uses a neural network based model that learns latent representations (concept embeddings) of clinical concepts based on how they appear in free text. The underlying algorithm is a modified version of the word2vec algorithm which learns numerical representations of a word based on the words that surround it.

Training MedCAT is done in two phases. The first is a self supervised phase in which MedCAT employs a simple technique to pre-annotate a large corpus of clinical text. In this step the algorithm identifies string matches for each concept synonym in the medical ontology being used, (e.g. searching for matches of "lung cancer" in each document). Once identified the word2vec algorithm is used to learn embeddings for those identified entities within the documents. This process provides MedCAT with an initial representation for how the concepts are represented in the free text.

In the second phase the model is fine-tuned using human provided annotations. In this case the model is taught to predict the correct label as provided by the human annotator using the MedCATtrainer interface (see 'Annotation Tools' section below). Based on some previous studies [12] the number of annotations required for fine tuning is small (500-600 annotated documents).

In addition to identifying clinical concepts in text MedCAT also provides a wrapper for training additional ML models for identifying important meta information for the extracted entities. Meta information of interest may include entity negated (e.g. 'patient does not have fever symptoms), if an identified entity relates to the patient or to somebody else (the experiencer), or whether it is current or historic. In order to implement these models, MedCAT uses a sequence based classifier (Bi-LSTMs) that takes the surrounding words of the identified terms and trains a classifier to predict if the meta label is assignable or not.

In summary, MedCAT is a machine learning model that is used to identify clinical concepts in clinical text and also reports important meta-information related to the extracted entities. At present, MedCAT is used to support individual clinical research projects but also more generally UCLH has a set of trained MedCAT models that it uses to annotate Epic data, sending these annotations back into other databases so that the annotations are accessible by the wider research community.  Additionally, MedCAT models have been shown to generalise well across multiple hospital settings with only minimal fine tuning required [12].

### Jupyter Hub and Building other ML Models:

Various clinical projects will require the development of custom machine learning models. We cover some of these in the case studies section. In order to facilitate the training of such models CogStack Nifi has been deployed with a JupyterHub instance. This provides data scientists, who want to work with data ingested into CogStack, with a platform for interacting with the data through Python.

### Annotation Tools:

Collecting annotated data for training machine learning models is done through a custom annotation interface. A custom set of interfaces was chosen over off the shelf ones (e.g. Doccano) as many of our annotation use cases require integrated tools for searching for clinical information.

MedCAT is trained using the MedCATtrainer interface [13]. The interface allows a user to load documents to be annotated by multiple annotators. The interface also provides an active learning mode that enables generated annotations to be used to retrain an existing MedCAT model in real time. The performance of the model can also be tracked in real time so the users can monitor performance change with additional annotations.

## Section 3: CogStack Use Cases:

CogStack services have been used to facilitate a number of clinical research projects.

### Clinical Trials Recruitment:

We used the CogStack platform in a retrospective simulation of patient recruitment into the LeoPARDS clinical trial [1], which studied a time-sensitive treatment for sepsis. We used NLP on free text clinical notes from the intensive care unit at UCLH to identify mentions of infection and medical diagnoses relevant to inclusion and exclusion criteria for the trial [2]. We then applied a rule based algorithm to identify eligible patients using a moving 1-hour time window, and compared patients identified by our approach with those actually screened and recruited for the trial.

Our method identified 376 patients, including all 34 patients with EHR data available who were actually recruited to LeoPARDS in our centre. The sensitivity of CogStack for identifying patients screened manually was 90% (95% CI 85%, 93%). Of the 203 patients identified by both manual screening and CogStack, the index date matched in 95 (47%) and CogStack was earlier in 94 (47%). We concluded that the CogStack platform with incorporated NLP could aid patient recruitment into a clinical trial, identifying some eligible patients earlier than manual screening, and could potentially improve trial recruitment by automatically identifying candidate patients if implemented in real time [2].

### NLP at the Point of Care:

UCLH has recently been involved in a national programme to develop an NLP system that can convert a clinician's text into structured information in real time, extracting information on diagnoses, medication and allergies. The new NLP system will communicate with the 'NoteReader' user interface component in Epic, which will allow clinicians to invoke the NLP system on their newly-created clinical note and generate the structured information, which can be verified before it is committed to the record. The current workflow for clinicians involves writing the clinical note and then proceeding to manually input information on diagnoses, comorbidities, medication and allergies into the appropriate structured fields.

The NLP system will use a trained MedCAT model, which will communicate with Epic NoteReader via a restful API. We have so far trained a MedCAT model on the entire UCLH record which includes clinical notes such as admission clerkings and discharge summaries. Specific training tests included patients with COVID-19 [14] and patients with heart failure, and in each case the model was trained to extract all diagnoses and symptoms, although for this project the output will be filtered to include only extracted concepts that clinicians would find useful to include on the problem list.

## Allergies

We used CogStack as part of a detailed analysis of adverse reaction reports submitted to the national National Reporting and Learning System (NRLS). The work focused on identifying reasons for why patients had an allergic reaction to prescribed and/or administered medications. The CogStack platform was used to collect annotations and train a multiclass classification model using sentence embeddings to identify a number of themes and causes that may have been involved, directly or indirectly, in the patient's adverse reaction.

The clinical collaborator, a consultant pharmacist, annotated a set of around 150 reports and labelled each report with one or more reasons for allergic reaction. A total of 20,788 incidents were extracted between 01/01/2012 and 31/12/2016. 6 key themes were identified including: time (night, out of hours); documentation (source, completeness, conflicts), knowledge (patient, medicine, cross-sensitivity), external or system factors (guidelines, microbiology advice/results, visual prompts), internal or individual factors (clinical condition, policy or procedure non-compliance, considered decision making) and medical/prescribing system (electronic or paper-based). 170 allergy reports were annotated and used to train the model.

The macro-f1 was 0.62 across all subthemes. The model reported higher f1s for simpler themes, such as temporary staff (1.0) and microbiology advice (0.93) whereas for more complex themes, such as non-compliance to policy (0.45), the reported f1s were lower. This was because unlike the simpler themes, the complex themes could not be identified through keywords/phrases, and the number of training examples in the dataset were too few for the model to be able to learn general semantic patterns for these themes.

## Improving Clinical Referrals Process for Neurology Clinics

CogStack is being used to analyse free text records for patients who had undergone a shunt insertion and/or lumbar drainage following a normal pressure hydrocephalus (NPH) diagnosis. In particular the clinicians are interested in developing an understanding of the most prevalent symptoms for patients undergoing these operations. A cohort study of patients attending the NPH clinic is being conducted on a sample of 500 annotated documents consisting of GP referral letters and imaging reports for a set of patients who had undergone the aforementioned procedures. The longer term objective of this project is to use the identified symptoms as features to train an ML model that can be used to flag potential patients who may have NPH and reduce any delays in referral to the clinic.

# Discussion/Conclusion

In this paper we have discussed UCLH's deployment of the low cost open source text analytics IR platform CogStack-Nifi. We have discussed the need for such a platform, namely the issues of ingesting data from multiple systems, the heterogeneity in data sources and most importantly text mining from the unstructured data. We discussed the key benefit of the updated CogStack-Nifi platform which is the ease in which dataflows can be set up and managed as well as the improved NLP capabilities of the platform. We also discussed the use of MedCAT in order to extract clinical concepts from the unstructured free text.

As demonstrated in our results section CogStack Nifi provides valuable systems and tools for supporting a wide range of clinical use cases. Consequently we feel that due to the low cost requirements of both the platform and the NLP models available with the platform, that CogStack Nifi can be deployed to most research focused healthcare organisations.